\documentclass[final, runningheads]{llncs}

\usepackage[T1]{fontenc}
\usepackage[belowskip=-10pt,aboveskip=7pt]{caption}

\usepackage{enumitem}
\usepackage[utf8]{inputenc}
\usepackage{hyperref} 
\usepackage{graphicx}
\usepackage{colortbl}
\usepackage[table,xcdraw]{xcolor}
\usepackage{ulem}
\usepackage{rotating}
\usepackage{cite}
\usepackage{multirow}
\usepackage{enumitem}
\setlist[itemize]{label=\textbullet}
\setlist[itemize]{label=$\circ$}

\usepackage{color}
\usepackage{tabularray}
\usepackage{xspace}

\usepackage{lipsum} \usepackage{tikz} \usepackage{enumerate}
\usepackage{graphicx,scalerel}
\usepackage{amsmath}

\usepackage{color}

\newcommand{\itadata}{\footnotesize \textsl{ITADATA2024: The 3$^{\text{rd}}$ Italian Conference on Big Data and Data Science}}
\usepackage{fancyhdr}
\fancypagestyle{empty}{\fancyhf{}\fancyhead[C]{\itadata}
  \fancyhead[R]{\includegraphics[width=.05\textwidth]{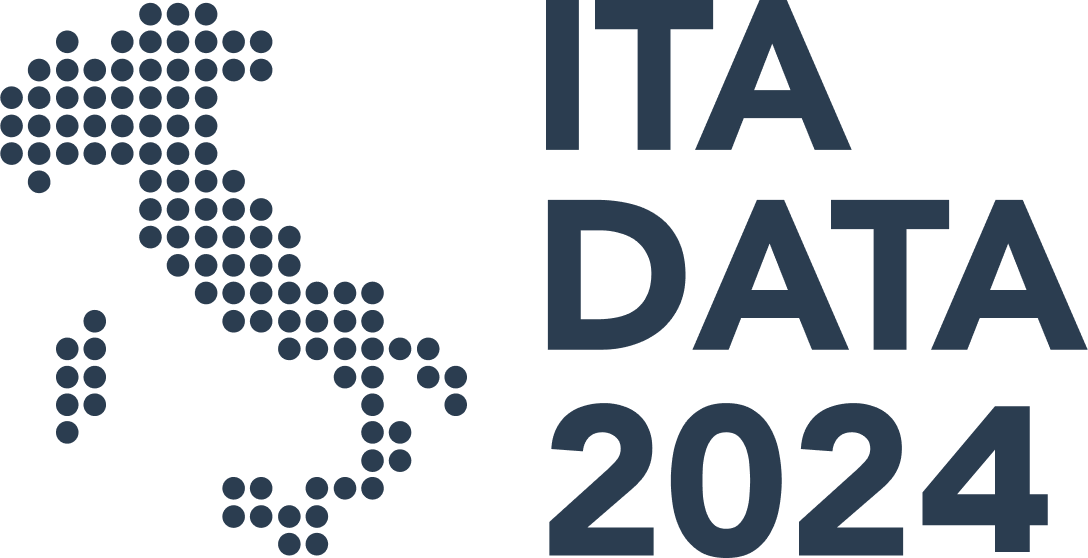}}}

\newcommand{\SoBigDataITAck}{European Union --- NextGenerationEU --- National Recovery and Resilience Plan (Piano Nazionale di Ripresa e Resilienza, PNRR) --- Project: ``SoBigData.it --- Strengthening the Italian RI for Social Mining and Big Data Analytics'' --- Prot. IR0000013 --- Avviso n. 3264 del 28/12/2021\xspace}

\makeatletter
\begin{document}

\title{An Evaluation Framework for the FAIR Assessment tools in Open Science}
\author{Payel Patra\inst{1}\orcidID{0000-0002-4615-4211} \and
Daniele Di Pompeo\inst{2}\orcidID{0000-0003-2041-7375} \and
Antinisca Di Marco\inst{3}\orcidID{0000-0001-7214-9945}}
\authorrunning{F. Author et al.}
\institute{University of L'Aquila, Italy
\\
\email{payel.patra@graduate.univaq.it, daniele.dipompeo@univaq.it, antinisca.dimarco@univaq.it
}
}

\maketitle 
\vspace{-1.4em}
\begin{abstract}
Open science represents a transformative research approach essential for enhancing sustainability and impact. Data generation encompasses various methods, from automated processes to human-driven inputs, creating a rich and diverse landscape. Embracing the FAIR principles—making data and, in general, artifacts (such as code, configurations, documentation, etc) findable, accessible, interoperable, and reusable— ensures research integrity, transparency, and reproducibility, and researchers enhance the efficiency and efficacy of their endeavors, driving scientific innovation and the advancement of knowledge.

Open Science Platforms OSP  (i.e., technologies that publish data in a way that they are findable, accessible, interoperable, and reusable) are based on open science guidelines and encourage accessibility, cooperation, and transparency in scientific research. Evaluating OSP will yield sufficient data and artifacts to enable better sharing and arrangement, stimulating more investigation and the development of new platforms.
 
In this paper, we propose an evaluation framework that results from evaluating twenty-two FAIR-a tools assessing the FAIR principles of OSP to identify differences, shortages, and possible efficiency improvements. 

\end{abstract}
\vspace{-1.3em}
\section{Introduction}\label{sec:intro}

Open science aims to provide access to scientific research and its dissemination to all facets of society~\cite{ref1,ref2}. Research artifacts are among the most valuable resources in open science since they serve as the basis for innovations that result in societal advantages like developing alternative energy sources or disease treatments. All digital resources should be able to be found, accessed, interoperable, and reused by machines and humans, according to the \href{https://www.gofair.org/}{FAIR Principles}.

For academics to publish scholarly results FAIRly must comply with the FAIR Principles\cite{ref3}, which serve as guidelines for the kinds of practices researchers and scientists should demand more and more from digital resources. The primary strategy is to keep the FAIR standards up-to-date across the global repositories. To achieve this goal, several tools help scientists make their artifacts (e.g., data, knowledge, code, etc) public by respecting the FAIR principles.

Since the FAIR principles are fundamental for Open Science\cite{ref4}, several platforms helping the researcher to publish their results are coming into action (e.g., \href{https://zenodo.org/}{Zenodo}, \href{https://sobigdata.d4science.org/}{SoBigData}, etc). In the following, we call such frameworks Open Science Platforms (OSP). Several FAIR assessment tools (called in the following FAIR-a tools) have been implemented to help scientists and researchers select the most appropriate OSP for their aims and scopes. Such tools assess how well the OSP frameworks implement the FAIR principles.

The main goal of this paper is to study FAIR-a tools and provide a means to evaluate and classify them to help the open science community select the best FAIR-a tool to be used to assess to what extent their publishing strategy responds to the FAIR principles. Our aim is to identify FAIR-a tools differences, shortages, and possible efficiency improvements.

In this paper we answer the following research questions:
\begin{itemize}
    \item \textbf{RQ1:} Which are the relevant features to be considered when assessing FAIR principles?
    \item  \textbf{RQ2:} Is the proposed evaluation framework suitable for classifying the existing FAIR-a tools?\item \textbf{RQ3:} Which FAIR-a tool performs better considering user perception? What are the primary shortages, and what are the possible efficiency improvements?
\end{itemize}

The contributions of this paper are:
\emph{i)} the specification of an evaluation framework for the FAIR-a tools useful to classify the existing FAIR-a tools. The evaluation framework has been defined using a bottom-up approach that extracts relevant features resulting from analyzing twenty-two FAIR-a tools linked at the \href{https://fairassist.org/}{webpage}
\footnote{\url{https://fairassist.org/}};   \emph{ii)} leveraging the defined evaluation framework, we provide an initial assessment of the existing FAIR-a tools; and finally, \emph{iii)} we report the list of best tools considering the user perspective.

In future work, we aim to enlarge the toolset to consider and involve more experts from science and industry in evaluating and improving the provided evaluation framework and its usage in the broader FAIR-a toolset.

\section{Research Motivation and Objectives}\label{sec:objective}

Numerous OSPs (that help open scientists to publish their dataset\cite{ref5} and artifacts following the FAIR principles) and assessment techniques (that assess to what extent the OSP adheres to the FAIR principles) have been published in earlier research \cite{ref7}. Due to the growing importance of open science, confirming the compliance of OSP to FAIR principles is now of paramount importance and requires a full investigation In particular, OSP aim to:
\begin{itemize}
\item ensure data integrity and trust, maintain high-quality, reliable repositories that meet Findable and Accessible standards. Enhancing data accessibility boosts research impact and encourages broader use of research findings, adhering to the Accessible principle.
\item Enhance data management effectiveness and efficiency by identifying and strengthening key areas. Promote stakeholder collaboration by implementing standardized data management planning (DMP) procedures, thus improving Interoperable and Reusable data practices.
\end{itemize}

Keeping these in mind, we can see how important it is to have FAIR assessment (FAIR-a) tools able to assess to what extent the OSP implements FAIR principles. However, FAIR-a tools currently have some shortcomings as well, therefore analyzing these tools might help identify the following points: 
\begin{itemize}
    \item It makes choosing the best tool easier. 
    \item Direct development and improvement of Tool's activities.
    \item It aids in the standardization of FAIR data practices.
    \item This will encourage developers to give FAIR principles priority.
    \item It will Organize research processes more efficiently. 
    \item It Provides transparency into tool's functionality and limitations.
\end{itemize}

In \cite{ref8}, N. Krans and the Karolinska Institutet research group (2022) evaluated ten FAIR assessment tools for chemicals and nanomaterials safety, noting inconsistencies in results and the need for consensus on assessment criteria. Tools like the ARDC FAIR self-assessment tool and FAIRshake offer various features, with SATIFYD recommended for quick assessments and automated tools aiding workflow development. Achieving consensus on FAIR criteria is vital for tool development. While online self-assessment tools suit individual datasets, (semi-) automated tools are better for entire databases.

Compared to \cite{ref8}, our study defines a novel evaluation framework with a bottom-up approach that evaluates 22 FAIR-a tools and provides detailed features and thorough assessment findings. Our comprehensive approach ensures superior performance and usability. The novelty and significance of our results within the FAIR context are underscored by the definition of an evaluation framework for FAIR-a tools, a detailed explanation of assessment results, and tool performance quality for these 22 tools. Moreover, this study goes beyond \cite{ref8} because we evaluated 22 FAIR-a tools while \cite{ref8} evaluated only 10 tools. This comprehensive assessment offers researchers a richer understanding and more precise differentiation of FAIR-a tools based on their qualities and capabilities, providing substantial insights for the research community.  
\section{Research Methodology}\label{sec:method}

\begin{figure}[!hbt]
    \centering
    \includegraphics[width=9.5cm, height=3.9cm]{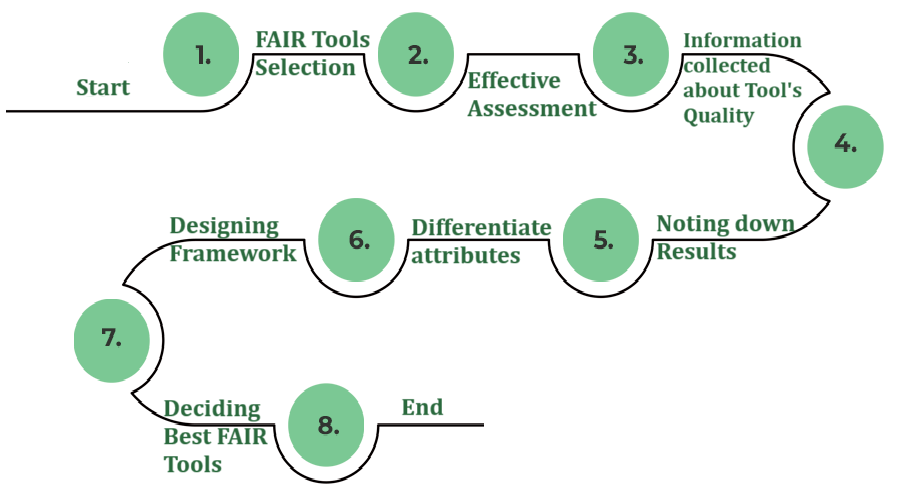}
\caption{FAIR-a tools evaluation methodology}
    \label{fig:Pic4}
\end{figure}

Figure \ref{fig:Pic4} sketches the methodology we followed in this study. It is composed by 8 steps starting with the selection of 22 FAIR-a 
tools publicly available.

Then, we conducted a thorough assessment of the considered toolset to ensure an effective evaluation process, yielding precise outcomes (step 2 in the figure). During the assessment, we note the dataset's qualities (step 3), facilitating the identification and differentiation of relevant attributes (steps 4 and 5) that formed the foundation for developing a robust framework, significantly contributing to our research field. 

Then, leveraging the identified tools' attributes and quality dimensions, we devised the evaluation framework (step 6) later used to evaluate the selected FAIR-a tools and identify the best ones (step 7) and close the methodology identifying potential research gaps (step 8).

In the assessment step, we used the SoBigData RI as OSP. SoBigData RI is particularly suitable for reviewing FAIR guidelines since its scope is to provide an open science platform to publish data and artifacts fully implementing FAIR principles.  Since many FAIR-a tools\cite{ref11} assess FAIR principle\cite{ref14} compliance of a specific open artifact, we considered the "Multilevel Monitoring of Activity and Sleep in Healthy People" paper and relative \href{https://physionet.org/content/mmash/1.0.0/}{dataset} from the SoBigData repository\cite{ref13}, released on June 19, 2020, version 1.0.0.

In Section \ref{sec:frame}, we report the relevant attributes identified in the assessment and describe the devised evaluation framework, while in Sections \ref{sec:func} and \ref{sec:result}
we report the preliminary results of the 22 FAIR-a tools' evaluation\cite{ref10}.

\section{Evaluation results}

\subsection{Addressing RQ1: Evaluation Framework Specification}\label{sec:frame}

Our proposed framework comprises 19 attributes grouped into 4 categories: functionality-related, technical-related, runtime-aspects, usability and user satisfaction. They are reported in Table \ref{tab4} and described in detail in the following.
The framework and the FAIR-a tools evaluation are available \href{https://github.com/PayelPatra/FAIR_project.git}{online}.

\begin{table*}[!ht]
\rotatebox{90}{
\resizebox{1.1\textwidth}{0.290\textheight}{

\begin{tabular}{|c|clllllllll|}
\hline
\rowcolor[HTML]{9B9B9B} 
\multicolumn{1}{|l|}{\cellcolor[HTML]{9B9B9B}}                                                                & \multicolumn{7}{c||}{\cellcolor[HTML]{9B9B9B}\textbf{Functionality-related}}   

& \multicolumn{3}{c||}{\cellcolor[HTML]{9B9B9B}\textbf{Technical-related}} 

\\ \hline
\rowcolor[HTML]{C0C0C0} 
\cellcolor[HTML]{9B9B9B}                                                            
 & \multicolumn{1}{c|}{\cellcolor[HTML]{C0C0C0}\textbf{\begin{tabular}[c]{@{}c@{}}Tool \\ Group\end{tabular}}} 

& \multicolumn{1}{c|}{\cellcolor[HTML]{C0C0C0}\textbf{\begin{tabular}[c]{@{}c@{}}Input \\ Types\end{tabular}}}                                                                                                                                                 & \multicolumn{1}{c|}{\cellcolor[HTML]{C0C0C0}\textbf{\begin{tabular}[c]{@{}c@{}}Target \\ Objects\end{tabular}}}

& \multicolumn{1}{c|}{\cellcolor[HTML]{C0C0C0}\textbf{\begin{tabular}[c]{@{}c@{}}Focused \\  Principle\end{tabular}}} 

& \multicolumn{1}{c|}{\cellcolor[HTML]{C0C0C0}\textbf{\begin{tabular}[c]{@{}c@{}}FAIR \\ Improvement \\ Scope\end{tabular}}} 

& \multicolumn{1}{c|}{\cellcolor[HTML]{C0C0C0}\textbf{\begin{tabular}[c]{@{}c@{}} Implemented \\ Strategy \end{tabular}}} 

& \multicolumn{1}{c||}{\cellcolor[HTML]{C0C0C0}\textbf{\begin{tabular}[c]{@{}c@{}}Result\\ Type\end{tabular}}}

& \multicolumn{1}{c|}{\cellcolor[HTML]{C0C0C0}\textbf{\begin{tabular}[c]{@{}c@{}}Degree \\ of Maturity\end{tabular}}} 

& \multicolumn{1}{c|}{\cellcolor[HTML]{C0C0C0}\textbf{\begin{tabular}[c]{@{}c@{}}Problems \\ To Fix\end{tabular}}} 

& \multicolumn{1}{c||}{\cellcolor[HTML]{C0C0C0}\textbf{\begin{tabular}[c]{@{}c@{}}Development \\ State\end{tabular}}}

\\ 

\cline{2-11} 

\multirow{-2}{*}{\cellcolor[HTML]{9B9B9B}\textbf{\begin{tabular}[c]{@{}c@{}}FAIR \\ Tool \\ Names\end{tabular}}} 

& \multicolumn{1}{l|}{\begin{tabular}[c]{@{}l@{}}* Online \\ Self-Assessment\\ *Online \\Semi-Automated\\ * Online \\ Automated\\ * Offline \\ Self-Assessmen\\ * Other Types \end{tabular}} 

& \multicolumn{1}{l|}{\begin{tabular}[c]{@{}l@{}}* Multiple\\ Choice\\ * Radio\\ * Input Box\\ * URL/PID/\\ DOI/GUID\\ * Digital \\ Objects\\ * Resource\\ * Software\\ * Repository\\ * Web \\ Questionnaire\\ * Manual \\ Checklist.\end{tabular}}

& \multicolumn{1}{l|}{\begin{tabular}[c]{@{}l@{}}* Dataset\\ * All Digital\\ Objects\\ * Terminology\\ Artifacts\\ * People's\\ Knowledge\\ * Software\end{tabular}} 

& \multicolumn{1}{l|}{\begin{tabular}[c]{@{}l@{}} *Findability \\ *Accessibility \\ *Interoperability \\ *Reusability \end{tabular}}                          

& \multicolumn{1}{l|}{\begin{tabular}[c]{@{}l@{}}* Yes plus \\ descriptive text\\ * No\end{tabular}}  

& \multicolumn{1}{l|}{\begin{tabular}[c]{@{}l@{}}* Assessment\\ * Improvement\end{tabular}}  

& \multicolumn{1}{l||}{\begin{tabular}[c]{@{}l@{}}* Qualitative\\ * Quantitative\end{tabular}}

& \multicolumn{1}{l|}{\begin{tabular}[c]{@{}l@{}}* Yes\\ * No\end{tabular}}      

& \multicolumn{1}{l|}{\begin{tabular}[c]{@{}l@{}}* Yes, plus \\ descriptive \\ text\\ * No\end{tabular}}

& \multicolumn{1}{l||}{\begin{tabular}[c]{@{}l@{}}* Developed\\ * Under \\ Development\end{tabular}}

\\ \hline

\\
\\

\hline
\rowcolor[HTML]{9B9B9B} 
\multicolumn{1}{|l|}{\cellcolor[HTML]{9B9B9B}}                                                                & \multicolumn{3}{c||}{\cellcolor[HTML]{9B9B9B}\textbf{Runtime aspects}}                           
& \multicolumn{7}{c||}{\cellcolor[HTML]{9B9B9B}\textbf{Usability and User Satisfaction}}

\\ \hline

\rowcolor[HTML]{C0C0C0} 
\cellcolor[HTML]{9B9B9B}                                                                                      

& \multicolumn{1}{c|}{\cellcolor[HTML]{C0C0C0}\textbf{\begin{tabular}[c]{@{}c@{}}Adaptability \\ (Extensibility)\end{tabular}}}

& \multicolumn{1}{c|}{\cellcolor[HTML]{C0C0C0}\textbf{\begin{tabular}[c]{@{}c@{}}Effort Cost \\ of Using Tool\end{tabular}}} 

& \multicolumn{1}{c||}{\cellcolor[HTML]{C0C0C0}\textbf{\begin{tabular}[c]{@{}c@{}}Execution\\ Time\end{tabular}}}

& \multicolumn{1}{c|}{\cellcolor[HTML]{C0C0C0}\textbf{\begin{tabular}[c]{@{}c@{}}Level of\\ Understanding\\ Question\end{tabular}}} 

& \multicolumn{1}{c|}{\cellcolor[HTML]{C0C0C0}\textbf{\begin{tabular}[c]{@{}c@{}}Level of \\ Understanding \\ Output\end{tabular}}}

& \multicolumn{1}{c|}{\cellcolor[HTML]{C0C0C0}\textbf{\begin{tabular}[c]{@{}c@{}}Level of\\ Tool's\\ Usability\end{tabular}}} 

& \multicolumn{1}{c|}{\cellcolor[HTML]{C0C0C0}\textbf{\begin{tabular}[c]{@{}c@{}}User's\\ Experience\end{tabular}}} 

& \multicolumn{1}{c|}{\cellcolor[HTML]{C0C0C0}\textbf{\begin{tabular}[c]{@{}c@{}}Recommend\\ To Others\end{tabular}}} 

& \multicolumn{2}{c||}{\cellcolor[HTML]{C0C0C0}\textbf{\begin{tabular}[c]{@{}c@{}}Overall \\ Evaluation \\ Grade\end{tabular}}}

\\ \cline{2-11} 

\multirow{-2}{*}{\cellcolor[HTML]{9B9B9B}\textbf{\begin{tabular}[c]{@{}c@{}}FAIR \\ Tool \\ Names\end{tabular}}} 

& \multicolumn{1}{l|}{\begin{tabular}[c]{@{}l@{}}* Yes\\ * No\end{tabular}}

& \multicolumn{1}{l|}{\begin{tabular}[c]{@{}l@{}}* Low\\ * Medium\\ * High\end{tabular}}

 & \multicolumn{1}{l||}{* Mins  }

& \multicolumn{1}{l|}{\begin{tabular}[c]{@{}l@{}}* Easy\\ * Medium\\ * Hard\end{tabular}}

& \multicolumn{1}{l|}{\begin{tabular}[c]{@{}l@{}}* Easy\\ * Medium\\ * Hard\end{tabular}}

& \multicolumn{1}{l|}{\begin{tabular}[c]{@{}l@{}}* Easy\\ * Medium\\ * Hard\end{tabular}} 

& \multicolumn{1}{l|}{\begin{tabular}[c]{@{}l@{}}* Best\\ * Good\\ * Medium\\ * NA\end{tabular}}                   

& \multicolumn{1}{l|}{\begin{tabular}[c]{@{}l@{}}* Yes\\ * No\\ * Partial\end{tabular}}

& \multicolumn{2}{c||}{\begin{tabular}[c]{@{}l@{}} *Good\\ * Better \\ * Best \end{tabular}}

\\ \hline
\end{tabular}}

}

\caption{The proposed Evaluation Framework}\label{tab4}
\vspace{-0.30cm}
\end{table*}

\noindent\textbf{Functionality-related Attributes.}

\begin{itemize}
\item  \textbf{Tool Group} attribute identifies the category the FAIR-a tool belongs to. In accordance with the definition given in \cite{ref15}, we consider 5 different groups:
\begin{itemize}
    \item  \underline{Online Self-Assessment} users can evaluate their datasets \textbf{online} through the survey. By simply visiting the website, they will find user-friendly \textbf{survey assessments}. After completing these surveys, insightful results will be received, either qualitative or quantitative, to enhance their understanding of their datasets. FAIR Data Self-Assessment Tool, FAIR-Aware, FAIRdat\cite{ref12}, FairDataBR, NFDI4Culture FAIR-Check, SATIFYD are under an online survey-based self-assessment group.
   \vspace{-0.7em}
    \item \underline{Online Semi-automated} tools use an automated procedure via a URL targeting all digital items. The tools ask questions with multiple-choice responses.
  
    \item \underline{Online Automated} tools ask users to provide  an online link, such as a DOI, PID, URL, resource link, or similar. After a while, the results of the assessment are displayed on the screen; it will take a few seconds to load up. 
    \item \underline{Offline Self-Assessment} group includes tools where users navigate the tool's website to download PDFs, Excel, or document files from the OSP they are published. Subsequently, users follow the provided instructions to complete a survey-type assessment, showcasing an innovative approach to evaluation. \item The last category encompasses \underline{Other types} of tools, offering awareness training through manual checklists and web questionnaires. \end{itemize}

\item The \textbf{Input type} attributes examine how tools are designed to receive user input for further assessments. The input type for evaluating FAIR-a tools varies, including \textit{multiple-choice questions, web-based questionnaires, and URL} inputs, depending on the tool's requirements and the nature of the evaluation.
\item \textbf{Target object} Once inputs are received, the tools \textbf{target} specific \textbf{objects} for subsequent assessments. These are \textit{Datasets, All digital objects, Terminology artifacts, People's knowledge, or Software}.
\item \textbf{Focused Principle} attributes identify which FAIR principle (Findability, Accessibility, Interoperability, and Reusability) the tool focuses on.
\item \textbf{FAIR Improvement Scope} This indicates if there is potential for future technical development to enhance the tool.
\item \textbf{Implemented Strategy} attribute, with its keyword "Assessment" and "Improvement", helps users understand whether the tool is used for assessing or improving FAIR principles.
\item \textbf{Result Type} attribute assesses the nature of the output (qualitative versus quantitative), with a preference for quantitative results.
\end{itemize}

\noindent\textbf{Technical-related Attributes.}

\begin{itemize}
    
 \item \textbf{Degree of Maturity} This shows whether the tool is still under or fully developed.
 \item Any technical issues that arise are mentioned in the \textbf{Problems to fix} components, highlighting areas needing attention from developers means \textit{(Yes or No)}. 
\item \textbf{Development State} attribute reflects the tool's guidance level and development status, indicating whether it is low/high on guidance or fully or partially developed (using the keywords "Developed" and "Under-Development"). 

 \end{itemize}

\noindent\textbf{Attributes related to Runtime Aspects}

\begin{itemize}   
    \item \textbf{Adaptability (Extensibility)} This indicates whether the tool is extensible by other users or developers, assessing its adaptability to other use case scenarios. 
    \item \textbf{Effort Cost of Using Tool} attribute represents the overall experience and effort required during the assessment. It can be Low, Medium or High.
    \item \textbf{Execution time} component records the duration of the entire assessment process, measured in \textit{(Minutes)}, to evaluate the tools' efficiency. In the case of not fully automated tools, this time also considers the time needed by the users to accomplish the tasks demanded of them.
\end{itemize} 

\noindent\textbf{Usability and User Satisfactory Attributes.}
\begin{itemize}   
    \item The \textbf{Level of understanding questions} component assesses the complexity of the questions posed by the tools, determining whether they are \textit{(Easy or Medium or Hard)} for users to understand. 
    \item Similarly, the \textbf{Level of understanding output} component measures how easily users can comprehend the results generated by the tools and mentioned with \textit{(Easy or Medium or Hard)}.
     \item The \textbf{Level of tool's usability} categorizes the tools based on their ease of use, ranging from easy to tough \textit{(Easy or Medium or Tough)}
    \item The overall \textbf{user experience} of FAIR tools is crucial for researchers. It varies based on interface design, instruction clarity, and resource availability. Tools with user-friendly interfaces, clear instructions, and ample resources offer the best experience. Those with moderately user-friendly interfaces and clear but sometimes challenging instructions provide a good experience. Tools with complex interfaces, unclear instructions, limited resources, and high manual effort present a medium experience, especially for users lacking advanced technical knowledge. These factors categorize the experience as \textit{(Best or Good or Medium or Not Applicable - NA)}.
    \item \textbf{Recommendations to others} are made based on the gathered information identifying the most effective and user-friendly tools and also determined by user feedback on their experience which are decided by \textit{Yes or Partial or No}.
    \item \textbf{Overall Evaluation Grade} attribute measures the overall quality of the tool, considering all aspects. It provides a subjective evaluation made by the user using \textit{Good, Better, or Best} values.
\end{itemize}

\paragraph{\textcolor{blue}{Answer to RQ1:}}  The proposed evaluation framework contains the main relevant features to be considered when assessing FAIR principles. They are grouped into 5 categories and cover very different aspects of a FAIR-a tool, including user satisfaction, functionality, and quality aspects of a FAIR assessment tool. If used in evaluating FAIR-a tools, the proposed framework helps to collect important features of existing tools that can guide towards the selection of the most suitable assessment tool or can highlights research gaps and inspire developers to create new, user-friendly FAIR tools with advanced techniques.

 \vspace{-0.9em}
\subsection{Addressing RQ2: using the evaluation framework to classify existing FAIR-a Tools}\label{sec:func}

Our research framework focuses on assessing the effectiveness and efficiency of FAIR-a Tools. We used it to assess the considered the 22 selected FAIR-a tools. We remark that online surveys have to be completed within 15 minutes to avoid \textit{survey fatigue} based on accepted survey methodological guidelines. Indeed, in our assessment session, we were not always able to complete a tool evaluation between 15 and 20 minutes due to poor user experience and the user-unfriendliness of the tool.  From the assessment, we obtained various insights; while certain tools encounter challenges due to technological limitations, others seamlessly integrate and perform well. This emphasizes the diverse capabilities and limitations of the assessed tools. In the following, we report insights from the evaluation of the 22 FAIR-a tools grouped by "Tool Group" attributes. For each group, we dedicate a section, and we report, for the sake of space, the evaluation of only one tool.
\vspace{-1.2em}
\subsubsection{Online Self-Assessment tools evaluation}

Tools in this group primarily use multiple-choice and radio options, sometimes combined with text boxes for specific criteria, to evaluate dataset quality and individuals' knowledge. Usability is judged by how easily the tools facilitate assessments, with documentation often provided for clarity. Tools are categorized as Easy, Medium, or Hard based on question clarity, UI design, and authentication procedures. All tools in this group are generally easy to handle, ensuring a comfortable and user-friendly experience. Output comprehension, both quantitative and qualitative, is crucial. Technical issues are also addressed to ensure a smooth user experience, rated as Good, Medium, or Best. All tools in this group completed the assessment within the 15-minute standard duration, promoting efficient research without compromising quality. Recommendations are based on user experience, with positive experiences leading to higher recommendations. FAIR Data Self-Assessment Tool, FAIR data Self-Assessment Tool, FAIR-Aware, FAIRdat, FairDataBR, NFDI4Culture FAIR-Check, SATIFYD belong to this group.

\paragraph{Assessment of SATIFYD tool:}
\textit{SATIFYD} is a user-friendly tool with drop-down, multiple-choice, and multi-select questions, featuring beautiful graphics. The evaluation took 10.58 minutes due to fewer questions. The results were Findable (88\%), Accessible (100\%), Interoperable (58\%), Reusable (85\%), and overall (83\%). The output page shows response values in blue within the letters F, A, I, and R, indicating a fair final score. User1 highly recommends the tool, citing ease of use, clear questions and outputs, and a positive experience with no issues.
\vspace{-1.4em}
\subsubsection{Online Semi-Automated tools evaluation}
 
Tools in this group use automated procedures via a URL for assessment, targeting all digital items. The tools ask questions with multiple-choice responses, but they are challenging to use and understand. Interpreting results is difficult, with no clear technological issues to address. Due to the poor user experience and inefficiency in completing the assessment within the specified timeframe, it is not recommended. FAIRshake tool is in this group.

\paragraph{Assessment of FAIRshake tool:}
To assess \textit{FAIRshake} we created a project named "SoBigData", specified the repository type, and used an existing rubric to progress the evaluation work. Despite spending 110 minutes understanding the tool and 20.73 minutes on evaluation, we found it challenging to grasp and questioned the sufficiency of self-evaluation for accurate FAIR results. Users need complete knowledge of all metrics to create their own, raising concerns about the tool's utility if users develop their metrics. Therefore, we do not recommend this tool unless necessary.

\vspace{-1.2em}
\subsubsection{Online Automated tools evaluation}
This group comprises nine FAIR tools, each with unique characteristics and various input types such as PID, URL, DOI, URI, software installation, digital objects, repositories, and other resources. These tools\cite{ref9} target digital objects, terminology artifacts, datasets, and software. Usability depends on clear documentation, intuitive UI design, and understanding the level of questions. Efficacy is measured by output accuracy, relevance, and meaningful responses. Performance varies, with some tools completing assessments within 15 minutes, while others take longer. Transparency about comprehension challenges is essential. These tools receive partial recommendations and need careful evaluation to address limitations. F-UJI, FAIR EVA (Evaluator, Validator \& Advisor), FAIR Evaluator, FAIR enough, FAIR-Checker, FOOPS!, HowFAIRIs, O'FAIRe, OpenAIRE Validator - FAIR assessment belong to this group.

\paragraph{F-UJI assessment:}
After completing the evaluation using the automated tool \textit{F-UJI}, it took 1 minute and 30 seconds to run, 7 minutes and 29 seconds to understand the output, totaling 8.59 seconds overall. The tool has an excellent output design with three steps: Evaluate Resource, Summary, and Report. It provided a pie chart showing a 62\% FAIR percentage. Scores were 'Advanced' in Findable [6/7], 'Moderate' in Accessible [3/4] and Interoperable [2/3], and 'Initial' in Reusable [2/3]. The detailed results were clear and easily understandable. We recommend this tool for future use.

\vspace{-1.4em}
\subsubsection{Offline Self-Assessment tools evaluation}
The Offline Self-Assessment tools offer an offline self-assessment similar to an easy-to-use survey, with questions ranging from moderate to medium complexity. Two standout tools in this category operate flawlessly without needing changes but exceed the 15-minute benchmark due to necessary processes and downloads. These assessments are crucial for understanding data FAIR principles. One tool is highly recommended for its effectiveness, while the other is somewhat recommended, highlighting the complex relationship between its usefulness and efficacy.
Do I-PASS for FAIR, FAIRness self-assessment grids belong to this group.

\paragraph{Assessment of "Do I-PASS for FAIR" tool:}
To evaluate  \textit{Do I-PASS for FAIR}, we downloaded the editable PDF from the "FAIR Enabling Research Organization" paper on November 3, 2020, and completed the evaluation in 22.05 minutes. The tool, consisting of 22 questions divided into Policy, Services, Skills, Incentives, and Adoption, rates answers as "Beginner," "Intermediate," or "Advanced." While some questions were irrelevant to non-administrators, the tool's usability and output were easy to understand. We can partially recommend it, as the overall experience was good.

\vspace{-1.2em}
\subsubsection{Other Types tools evaluation}
The fifth and final group, categorized under "other types", comprises three tools targeting all digital objects. These tools use web and manual questionnaires, along with manual checklists, as input types. Usability levels vary, and some assessments remain incomplete. However, all tools in this group are recommended due to their training and lessons. Despite their effectiveness, these tools exceed the 15-minute benchmark for completing both the assessment and procedural steps. Data Stewardship Wizard, TRIPLE Training Toolkit, and FAIRplus are under the fifth group.

\paragraph{Assessment of FAIR-DSM tool:}
The FAIR DataSet Maturity (FAIR-DSM) Assessment Tool, part of the European project  \textit{\textbf{FAIRplus}}, has 17 questions covering five maturity levels with various indicators. The assessment took 15 minutes and 22 seconds with medium comprehension checklist selections. The tool includes guidelines and materials from the FAIR COOKBOOK to help make data FAIR. The output page detailing all maturity indicators was easy to understand. Overall, user experience was medium, and we recommend this tool.

\paragraph{\textcolor{blue}{Answer to RQ2:}} The comprehensive nature of our framework allows for a detailed analysis of each tool's performance, ensuring that the best, most time-efficient FAIR-a tools are selected. By providing a thorough overview, our framework reduces the effort needed for selection, promoting data handling best practices and enhancing tool development accuracy. It guides data management planning and supports data privacy in the medical field, ensuring robust methods for testing FAIR-a tools' performance. Additionally, our framework makes data more reproducible, valuable, and ready for future reuse, empowering researchers, developers, and data managers to optimize workflows and achieve better outcomes. This comprehensive approach ensures that the chosen FAIR-a tools are effective and efficient and align with the latest advancements and requirements in the field, making it a superior choice compared to the standard approach.

 \vspace{-1.2em}
\subsection{Addressing RQ3: identify the best FAIR-a tool
from the user perspective}\label{sec:result}

After evaluating the FAIR-a tools, we assessed their quality and conducted a comparative analysis, organizing them into a table based on the original research framework. For instance, we analyzed the FAIR-a tool \textbf{SATIFYD} by checking its attributes. 'SATIFYD' is in a developed state, and lacks adaptability, but has a certain degree of maturity, providing quantitative results. It focuses on the four FAIR principles, emphasizes the FAIR Assessment strategy, and requires low effort to use. Based on user evaluations, 'SATIFYD' is considered one of the best tools, though improvements like increasing knowledge and verifying FAIR data are needed.

\begin{figure}[!hbt]
  \centering
 \includegraphics[width=9.5cm, height=6.2cm]{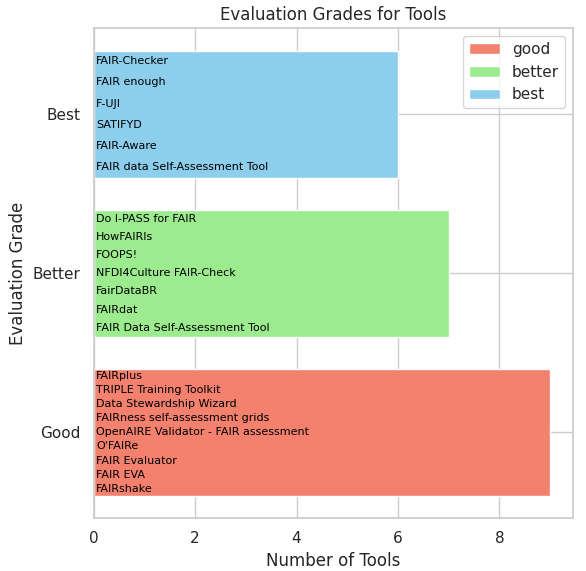}
   \caption{This figure represents Evaluation Grade as bars with FAIR-a Tool Names.}
    \label{fig:Pic2}
\end{figure}

Through this framework, we have identified the top six FAIR-a tools that excel in every dimension: FAIR-Checker, FAIR enough\cite{ref6}, F-Uji, SATIFYD, and FAIR-Aware, FAIR data Self-Assessment Tool categorized by specified bars, as illustrated in Figure \ref{fig:Pic2}. Additionally, there are seven tools that, while not the best in all actions, are still good enough and provide substantial value: Do I-PASS for FAIR, HowFAIRIs, FOOPS!, NFDI4Culture FAIR-Check, FairDataBR, FAIRdat, and FAIR Data Self-Assessment Tool.

Despite the many advantages, several challenges and limitations exist when working with the tools in our framework: Understanding suitable data types for assessment and ensuring compatibility with different data formats is a primary concern. Some tools require standalone implementation, which can be hectic and complex. Additionally, although guidance is provided on platforms like GitHub, individuals without extensive computer knowledge may struggle to understand the entire procedure.

\vspace{-1em}
\paragraph{\textcolor{blue}{Answer to RQ3:}} Through our framework, we were able to identify the overall better FAIR-a tools. In particular, 6 out of the 22 evaluated showed the highest level of user satisfaction. Moreover, we were able to identify the main challenges and limitations of using the tools.
\vspace{-1em}
\section{Threats to validity}

The evaluation framework has been tested using the \textit{"Multilevel Monitoring of Activity and Sleep in Healthy People (MMASH)"} dataset from the SoBigData repository. Even if it is a particular dataset, the results and effectiveness of the FAIR-a tools are not affected by the choice. The generalizability and robustness of the study results can be affected by the 22 FAIR-a tools considered to construct the evaluation framework and by the fact that we used SoBigData RI as OSP. In future work, to validate the evaluation framework, we will extend the FAIR-a toolset and run the experiments using other OSP (such as ZENODO).

A single researcher has conducted the evaluation of the 22 FAIR-a tools. To make the evaluation of the FAIR-a tools more robust and general, we plan to include more open science experts in the study.

 \vspace{-0.30cm}
\section{Conclusion And Future Work}\label{sec:conclusion}
Our research focused on evaluating FAIR assessment tools and on the definition of an evaluation framework using a bottom-up approach. The proposed framework permitted the classification of the 22 FAIR-a tools and the selection of the top ones, also aiding new researchers by providing clear evaluation criteria and detailed FAIR scores. It also serves as a valuable resource for developers, highlighting areas needing improvement.

Future work includes validating findings through additional evaluations and developing a new FAIR-a tool to address identified drawbacks such as incompatibility with different data formats. This aims to enhance global open science artifacts management practices, making them more reproducible, valuable, and ready for future reuse, ultimately contributing to the ongoing improvement of data management practices worldwide.

\vspace{-0.30cm}
\begin{credits}
\subsubsection*{\ackname} This work is supported by \SoBigDataITAck. 
\end{credits}

\end{document}